\documentclass[12pt]{article}

\usepackage{latexsym} 
\usepackage{amssymb}  
\usepackage{epsfig}       

\newcommand{\eps}{\epsilon}

\newcommand{\beq}{\begin{equation}}
\newcommand{\eeq}{\end{equation}}
\newcommand{\ba}{\begin{array}}
\newcommand{\bea}{\begin{eqnarray}}
\newcommand{\ea}{\end{array}}
\newcommand{\eea}{\end{eqnarray}}

\newcommand\comment[1]{ \hbox{[{\it Comment suppressed here.}\/]} }
\newcommand\hide[1]{}

\newcommand{\Tr}{\hbox{Tr}}

\newcommand{\bp}{{\bf p}}

\newcommand{\skipover}[1]{}

\newcommand{\addup}{\texttt{addup} }
\newcommand{\fft}{\texttt{fft} }
\newcommand{\invfft}{\texttt{invfft} }

\def\figput#1{\epsfig{file=#1,width=10cm}}
\makeatletter 

\def\appendix{\par                              
    \setcounter{section}{0}                     
    \setcounter{subsection}{0}
    \renewcommand{\theequation}{\Alph{section}.\arabic{equation}}
    \renewcommand{\thesection}{Appendix \Alph{section}}
}

\def\applabel#1{\@bsphack
  \protected@write\@auxout{}%
         {\string\newlabel{#1}{{\Alph{section}}{\thepage}}}%
  \@esphack}

\def\section{
\setcounter{equation}{0}        
\@startsection {section}{1}{\z@}{-3.5ex plus -1ex minus
 -.2ex}{2.3ex plus .2ex}{\large\bf}}
\renewcommand{\theequation}{\arabic{section}.\arabic{equation}}

\def\subsection{\@startsection{subsection}{2}{\z@}{-3.25ex plus -1ex minus
 -.2ex}{1.5ex plus .2ex}{\normalsize\bf}}

\def\subsubsection{\@startsection{subsubsection}{3}{\z@}{-3.25ex plus
 -1ex minus -.2ex}{1.5ex plus .2ex}{\normalsize}}

\makeatother   

\newsavebox{\eqlabel}

\makeatletter  
\newlength{\numblen}
\newsavebox{\eqnumb}
\def\@eqnnum{\savebox{\eqnumb}{\rm (\theequation)}%
\settowidth{\numblen}{\usebox{\eqnumb}}%
\makebox[\numblen][l]{\usebox{\eqnumb}~~~\usebox{\eqlabel}}}
\makeatother   

\newenvironment{equationwithlabel}[1]{ %
  \savebox{\eqlabel}{#1}
  \begin{equation}\label{#1} }{\end{equation}} 
\newcommand{\beql}[1]{\begin{equationwithlabel}{#1}}
\newcommand{\eeql}{\end{equationwithlabel}}

\begin{document}

\title{\bf Renormalized thermodynamics from the
2PI effective action
\\[1.ex]}

\author{
J.~Berges\thanks{email: j.berges@thphys.uni-heidelberg.de} $\,^a$,
Sz.~Bors\'anyi\thanks{email: s.borsanyi@thphys.uni-heidelberg.de} $\,^a$,
U.~Reinosa\thanks{email: reinosa@hep.itp.tuwien.ac.at} $\,^{ab}$,
and J.~Serreau\thanks{email: j.serreau@thphys.uni-heidelberg.de} $\,^a$
\\[1.ex]
{$^a$ Universit\"at Heidelberg, Institut f\"ur
Theoretische Physik}\\
{Philosophenweg 16, 69120 Heidelberg, Germany}\\
{$^b$ Institut f\"ur Theoretische Physik, Technische Universit\"at Wien}\\
{Wiedner Hauptstrasse 8-10/136, A-1040 Wien, Austria}
}

\date{}

\begin{titlepage}
\maketitle
\def\thepage{}          

\begin{abstract}
\noindent
High-temperature resummed perturbation theory is plagued by poor
convergence properties. The problem appears for theories with bosonic 
field content such as QCD, QED or scalar theories.
We calculate the pressure as well as other thermodynamic
quantities at high temperature for a scalar one-component field
theory, solving a three-loop 2PI effective action numerically without
further approximations. We present a detailed comparison
with the two-loop approximation. One observes a strongly improved
convergence behavior as compared to perturbative approaches.
The renormalization employed in this work extends previous prescriptions,
and is sufficient to determine all counterterms required for the theory
in the symmetric as well as the spontaneously broken phase.
\end{abstract}

\end{titlepage}

\renewcommand{\thepage}{\arabic{page}}


\section{Introduction and overview}
\label{sect:intro}

All information about the quantum theory can be obtained from the effective
action, which is the generating functional
for Green's functions. Typically, the (1PI)
effective action $\Gamma[\phi]$ is represented as a functional of the
field expectation value or one-point function $\phi$ only. In contrast,
the so-called two-particle irreducible (2PI)
effective action $\Gamma[\phi,D]$ is written as a functional
of $\phi$ and the connected two-point
function $D$~\cite{Baym}.
The latter provides an efficient description
of quantum corrections in terms of resummed loop-diagrams.
The different functional representations of the
effective action are equivalent in the sense that they are
generating functionals for Green's functions including all
quantum/statistical fluctuations and agree by construction in the absence of
sources. However, e.g.~loop expansions
of the 1PI effective action to a given order in the presence of
the ``background'' field $\phi$ differ in general from a loop expansion
of $\Gamma[\phi,D]$ in the presence of $\phi$ and $D$.

This observation has been successfully used in nonequilibrium
quantum field theory to resolve the problems of secularity and late-time
universality~\cite{Berges:2000ur,Berges:2001fi}
of perturbative approximations, which render the latter
invalid even for arbitrarily small couplings \cite{reviews}. Both the
far-from-equilibrium behavior as well as the \mbox{late-time}
thermal equilibrium results can be described from a loop expansion
of the 2PI effective action\footnote{Loop
approximations of the 2PI effective
action are also called ``$\Phi$-derivable''.}
without further assumptions for scalar~\cite{Berges:2000ur,Aarts:2001qa}
and fermionic theories~\cite{Berges:2002wr}. Similar results
have been obtained using a two-particle irreducible $1/N$ expansion
beyond leading
order~\cite{Berges:2001fi,Aarts:2001yn,Cooper:2002qd}.

The same techniques can be applied directly in thermal equilibrium,
where efficient formulations in Euclidean space-time become available.
In contrast to the far-from-equilibrium case, there are various powerful
approximation schemes known in thermal field theory.
A prominent approach in equilibrium high-temperature
field theory is the so-called ``hard-thermal-loop''
resummation~\cite{Braaten:1989mz}. However, explicit
calculations of thermodynamic quantities such as pressure
or entropy typically reveal a poor
convergence except for extremely small couplings. An important example
for this behavior concerns high-temperature gauge theories. 
Recent strong efforts to improve the convergence aim at connecting 
to available lattice QCD results, for which high temperatures
are difficult to achieve. In order to find
improved approximation schemes it is important to note that the
problem is not specific to gauge field theories. Indeed
it has been documented in the
literature in great detail that problems of convergence of perturbative
approaches at high temperature can already be studied in simple scalar
theories. For recent reviews in this context see Ref.~\cite{Blaizot:2003tw}.

A promising candidate for an improved convergence behavior is the loop or
coupling expansion of the 2PI effective action. So far, thermodynamic
quantities such as pressure or entropy have been mainly calculated to two-loop
order. However, aspects of convergence can be sensefully discussed only beyond
two-loop order since the one-loop high-temperature result corresponds
to the free gas approximation.
Efforts to calculate pressure beyond two loops include
so-called approximately self-consistent
approximations~\cite{Blaizot:1999ip}\footnote{Cf.~also
Ref.~\cite{Gruter:2004bb} for a similar strategy in the context of
Schwinger-Dyson equations.},
as well as estimates based on further perturbative expansions in the
coupling and a variational mass
parameter~\cite{Braaten:2001vr}\footnote{Cf.~Ref.~\cite{Andersen:2004re} for a
similar application to QED.}. These studies
indicate already improved convergence
properties. However, perturbatively motivated estimates as in
Ref.~\cite{Braaten:2001vr} suffer from the presence of nonrenormalizable,
ultraviolet divergent contributions and the apparent breakdown of the approach
beyond some value for the coupling.
If one does not want to rely on these further assumptions, going
beyond two-loop order requires the use of efficient numerical techniques. Such
rigorous studies are important to get a decisive answer about the properties of
2PI expansions. As it turns out (cf.~below) these problems
appear as an artefact of the additional approximations employed and cannot be
attributed to the 2PI loop expansion.

\begin{figure}[t]
\begin{center}
\epsfig{file=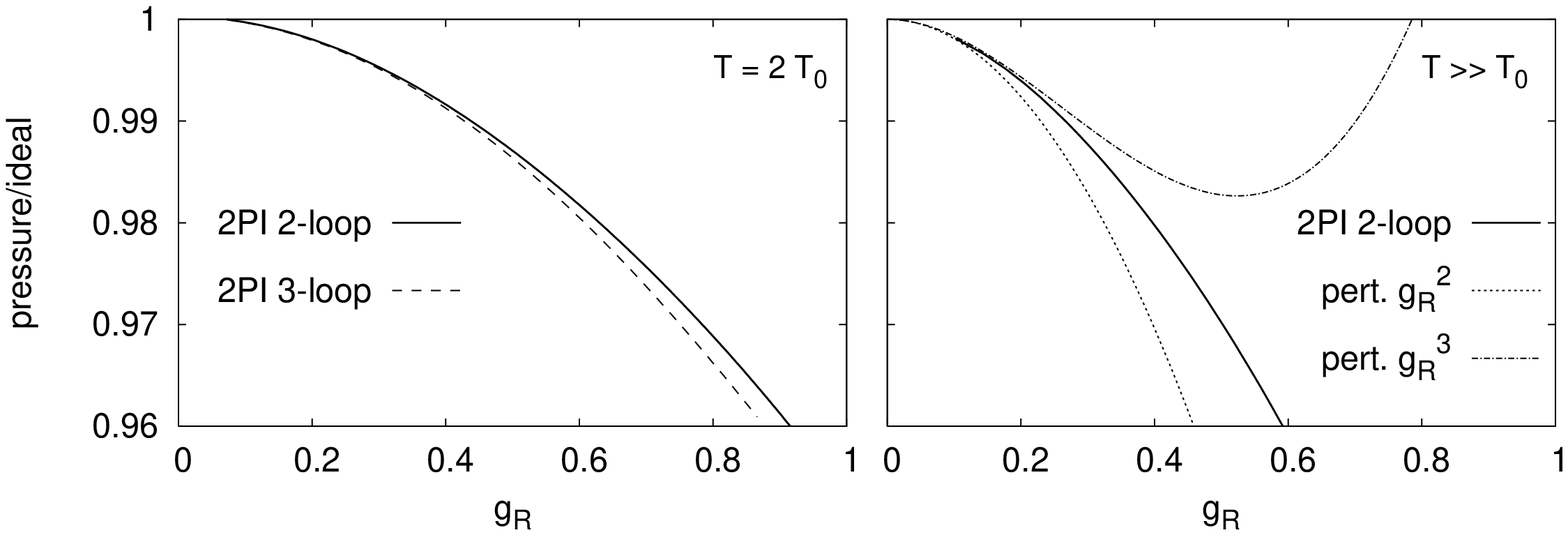,width=16.5cm,height=6.5cm}
\end{center}
\vspace*{-0.5cm}
\caption{Pressure as a function of the renormalized coupling,
normalized to the ideal gas, i.e.~one-loop pressure. Shown are the 2PI
two- and three-loop results (left) for $T = 2T_0$ with $T_0=m_R(T_0)$.
The right figure shows the perturbative results to order $g_R^2$ and $g_R^3$
along with the 2PI two-loop curve in the high-temperature limit for
illustration of the problematic alternating behavior of perturbation
theory (see text).
}
\label{fig:pressure}
\end{figure}
In this work we calculate the pressure as well as other
thermal quantities for a scalar $g^2 \phi^4$ field
theory from a three-loop 2PI effective action numerically without
further approximations.  A detailed comparison
with the two-loop approximation is presented. We observe a strongly improved
convergence behavior as compared to perturbative approaches. This is
exemplified in Fig.~\ref{fig:pressure}, where the pressure is shown
as a function of the renormalized coupling $g_R$ determined by the physical
four-vertex. The left figure compares the two- and
three-loop result normalized to the ideal gas pressure. For the employed
high temperature $T = 2 m_R(T_0)$ the three-loop corrections to the
pressure are rather small. Here $m_R(T_0)$ is the temperature-dependent
renormalized mass parameter or inverse correlation length and we
have $T_0 = m_R(T_0)$.
For illustration we also show on the right of Fig.~\ref{fig:pressure}
the perturbative results to order
$g_R^2$ and $g_R^3$ along with the dominant 2PI two-loop result
for the high-temperature limit.
The problematic alternating behavior of the perturbative series
is not specific to the limit $T \gg T_0$ and is characteristic
for higher orders as well~\cite{fiveloop}. (Cf.~Secs.~\ref{sec:analytic}
and~\ref{sec:numerics_nlo}.)

We obtain the renormalized correlation functions or proper
vertices as functions of temperature, building on a renormalization put forward
in Refs.~\cite{HeesKnoll,BIR} (cf.~also \cite{Cooper:2004rs}).\footnote{For
related studies see also Ref.~\cite{Jakovac:2004ua}.
In contrast to the considered
``non-local'' 2PI resummation, these investigate renormalization for
``local'' resummations. The latter turn out to be problematic to describe
the nonequilibrium late-time behavior of quantum fields~\cite{Baacke:2003qh}
and will not be discussed here.}
In contrast to previous approaches, the renormalization employed in
this work is formulated for the resummed 1PI effective action which
is calculated from a given loop approximation of the 2PI effective action.
The procedure is sufficient to determine all counterterms
required for the theory in the symmetric as well as spontaneously broken
phase. This extends previous
prescriptions, which are not sufficiently general
to renormalize the theory in the presence of a non-vanishing field.
In particular, they do not renormalize all functional field-derivatives
of the effective action even in the symmetric phase. The latter represent
the proper vertices, which encode important information about the theory.
This is discussed in Sec.~\ref{sec:renormalization} taking into account
three-loop corrections. The considered approximation represents a
simple explicit example of a systematic renormalization scheme for
2PI effective actions, which will be given for general approximations
in a separate publication~\cite{BBRS2}.

To put our calculations in a more general context, we note that our
results for the two- and three-loop approximations of the 2PI
effective action are identical to the corresponding two- and
three-loop approximations for $n$-particle irreducible ($n$PI)
effective actions with arbitrary $n \ge 3$. The agreement up to the
considered order is a consequence of an equivalence hierarchy for
$n$PI effective actions~\cite{Berges:2004pu}. Their functional
dependence takes into account the propagator as well as the proper
three-vertex, four-vertex, \ldots,
$n$-vertex~\cite{Baym,Dominicis,Calzetta:1988cq,Berges:2004pu}.
Therefore, a loop expansion of the $n$PI effective action with $n
> 2$ treats propagators and the respective vertices on the same footing, while a 2PI
loop expansion singles out propagator resummation a priori. However,
the difference to the 2PI results in the symmetric phase appears
only at four-loop order~\cite{Berges:2004pu}, which is beyond the
approximation employed in this work. As a consequence, the good
convergence properties of the expansion, which we observe, may be
attributed to the $n$PI loop expansion with arbitrary $n > 2$ rather
than to the 2PI loop expansion.\footnote{\label{foot:1} An
indication for this is that the quantitative description of the
universal behavior near the second-order phase transition of this
model goes beyond a 2PI loop expansion~\cite{Alford:2004jj}. It
requires taking into account vertex corrections that start with the
4PI effective action to four-loop order. The latter agrees with the
most general $n$PI loop expansion to that
order~\cite{Berges:2004pu}.}

\section{Renormalized thermodynamics}
\label{section2pi}

We consider a quantum field theory with classical action\footnote{We
employ a Minkowskian metric $g^{\mu\nu} = \mbox{diag}(1,-1,-1,-1)$
in view of further possible applications, e.g.~in out-of-equilibrium
situations.}
\begin{equation}
S = \int {\rm d}^4 x \left\{ \frac{1}{2} \left(
\partial_{\mu} \phi
\partial^{\mu} \phi
- m^2 \phi^2 \right)
- g^2 \phi^4 \right\} \, ,
\label{eq:classical}
\end{equation}
where $\phi(x)$ is a real
scalar field with bare mass term $m^2$
and coupling $g^2$. The normalization of the
coupling is chosen for simple comparison with existing literature
(cf.~e.g.~Ref.~\cite{Blaizot:1999ip}) in view of applications of these
methods to QCD thermodynamics. We use the shorthand
notation $\int_x \equiv \int_{0}^{- i/T} {\rm d} x^0
\int {\rm d}^3 x$ with temperature $T$.
Following~\cite{Baym} it is convenient to parametrize the
temperature dependent 2PI
effective action as
\begin{equation}
\Gamma[\phi,D] = S[\phi] + \frac{i}{2} \Tr\ln D^{-1}
          + \frac{i}{2} \Tr D_0^{-1}(\phi) D
          + \Gamma_2[\phi,D] +{\rm const}\, ,
\label{eq:2PIaction}
\end{equation}
which expresses $\Gamma$ in terms of the classical action $S$ and correction
terms including the function $\Gamma_2$ to which only two-particle irreducible
diagrams contribute.  Here the classical inverse propagator is given by
$i D_{0}^{-1}(x,y;\phi) \equiv
\delta^2 S[\phi]/\delta \phi(x) \delta \phi(y)$.
In the absence of external sources physical solutions require
\begin{eqnarray}
\frac{\delta \Gamma[\phi,D]}{\delta \phi(x)}\Big|_{\phi = \phi_0} &=& 0 \, ,
\label{eq:phistationary}\\
\frac{\delta \Gamma[\phi,D]}{\delta D(x,y)}\Big|_{D = D(\phi)}
&=& 0 \, .
\label{eq:stationary}
\end{eqnarray}
The 2PI effective action evaluated at $D(\phi;x,y)$,
i.e. for the $\phi$-dependent solution of (\ref{eq:stationary}),
is identical to the 1PI effective action $\Gamma[\phi,D(\phi)]$.
The effective action at the stationary point, $\Gamma[\phi_0,D(\phi_0)]$,
corresponds to the logarithm of the partition function in the absence of
sources~\cite{Baym}.  Therefore, in thermal equilibrium with
temperature $T$ ($\phi_0$ constant) the effective action is related to
the pressure $P$ by
\begin{equation}
P = \frac{T}{L_3} i \Gamma[\phi_0,D(\phi_0)]\, ,
\label{eq:relpg}
\end{equation}
where $L_3 = \int {\rm d}^3 x$ denotes the spatial volume and the constant in
Eq.~(\ref{eq:2PIaction}) is chosen such that the pressure vanishes at zero
temperature. Entropy density ${\mathcal S}$ and energy density ${\mathcal E}$
are given by
\begin{eqnarray}
{\mathcal S} &=& \frac{\partial P}{\partial T} \, ,\\
{\mathcal E} &=& - P + T {\mathcal S} = T^2 \frac{\partial}{\partial T}
\left( \frac{P}{T} \right) \, .
\end{eqnarray}
We recall that all the physical information is contained in the
effective action at the stationary point $\Gamma[\phi_0,D(\phi_0)]$
and its changes with respect to variations in the field $\phi$
evaluated at $\phi = \phi_0$.
For instance, the connected two-point function, $\Gamma^{(2)}$, and proper
four-point function, $\Gamma^{(4)}$, are given by
\bea
\Gamma^{(2)}(x,y) &\equiv&
\frac{\delta^2 \Gamma[\phi,D(\phi)]}{\delta \phi(x) \delta \phi(y)}
\Big|_{\phi=\phi_0} \, ,
\label{eq:2point}
\\
\Gamma^{(4)}(x,y,z,w) &\equiv&
\frac{\delta^4 \Gamma[\phi,D(\phi)]}{\delta \phi(x)
\delta \phi(y) \delta \phi(z) \delta \phi(w)}
\Big|_{\phi=\phi_0}\, .
\label{eq:4point}
\eea
All information about the quantum theory can therefore
be conveniently obtained from $\Gamma[\phi,D(\phi)]$
by functional differentiation. In particular, $\Gamma[\phi,D(\phi)]$
evaluated for constant field $\phi$ encodes the effective potential.

\subsection{Renormalization}
\label{sec:renormalization}

The effective action $\Gamma[\phi,D(\phi)]$
is defined in a standard way by suitable
regularization, as e.g.~lattice regularization or
dimensional regularization, and renormalization conditions.
We employ renormalization conditions for the
two-point function (\ref{eq:2point}) and four-point
function (\ref{eq:4point}), which in Fourier space
read:
\bea
Z\, \Gamma^{(2)}(p^2)|_{p = 0} &=& - m_R^2 \, ,
\label{eq:r1}\\
Z\, \frac{d}{d p^2} \Gamma^{(2)}(p^2)|_{p = 0} &=& - 1 \, ,
\label{eq:r2}\\
Z^2\, \Gamma^{(4)}(p_1,p_2,p_3)|_{p_1 = p_2 = p_3 = 0} &=&
- 4!\, g^2_R \, ,
\label{eq:r3}
\eea
with the wave function renormalization $Z$. Here the renormalized
mass parameter $m_R$ corresponds to the inverse correlation
length. The physical four-vertex at zero momentum is given by $g^2_R$.
Without loss of generality
we use throughout this work renormalization conditions for $\phi_0 = 0$.

We emphasize that at non-zero temperature the mass parameter as well as the
four-vertex are temperature dependent. The value of the mass
and coupling at a given temperature and momentum scale uniquely
determines the theory. This can then be used to calculate
properties at some other temperature.
We will typically define the theory by giving renormalization conditions
at zero temperature as well as some non-zero temperature $T_0$ with
$m_R \equiv m_R(T_0)$ and $g_R \equiv g_R(T_0)$.

\subsubsection{2PI renormalization scheme to order $g_R^4$}
\label{ssec:scheme}

In order to impose the renormalization conditions
(\ref{eq:r1})--(\ref{eq:r3}) one first has to calculate
the solution for the two-point field $D(\phi)$ for $\phi = 0$,
which encodes the resummation and which is obtained from the stationarity
condition for the 2PI effective action (\ref{eq:stationary}).
To achieve this we will for the most part follow the lines of the
renormalization procedure described in Ref.~\cite{BIR}.
We comment on the additional implications, which arise from imposing
(\ref{eq:r1})--(\ref{eq:r3}), below. For further details we refer to
Ref.~\cite{BBRS2}. Here we will explicitly demonstrate
from the numerical solution of the three-loop approximation that the
renormalized quantities are insensitive to the change of the (lattice)
regularization.

\begin{figure}[t]
\begin{center}
\epsfig{file=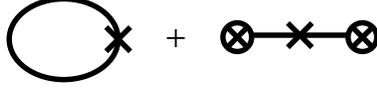,width=5.cm}
\caption{\label{fig:ct} Diagrammatic representation of the mass and field
strength counterterms. The cross in the left graph denotes indistinctly
the mass counterterm $\delta m_1^2$ or field strength counterterm
$\delta Z_1^2$ appearing in Eq.~(\ref{eq:counter1loop}), while the one in the
right graph represents the $\delta m^2$ and $\delta Z^2$ insertions
in Eq.~(\ref{eq:counterS}). The line of the closed loop represents
$D_R$, while a circled cross denotes $\phi_R$.}
\end{center}
\end{figure}

\begin{figure}[t]
\begin{center}
\epsfig{file=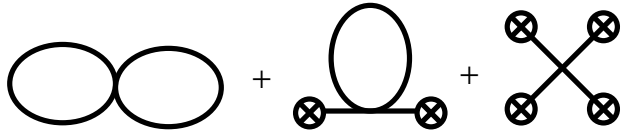,width=8.2cm}
\caption{\label{fig:tadpoles} Diagrammatic representation of
the coupling counterterms. Here the vertices denote
$\delta g_2^2$, $\delta g_1^2$ and
$\delta g^2$, respectively, which appear in Eqs.~(\ref{eq:G2orderg4}),
(\ref{eq:counter1loop}) and (\ref{eq:counterS}).}
\end{center}
\end{figure}

The renormalized field is $\phi_R = Z^{-1/2} \phi$. It is
convenient to introduce the counterterms relating the bare
and renormalized variables in a standard way with
\begin{eqnarray}
Z m^2 = m_R^2 + \delta m^2
\,\, , \qquad
Z^2 g^2 = g^2_R + \delta g^2
\,\, , \qquad \delta Z = Z - 1 \, ,
\label{eq:introcount}
\end{eqnarray}
and we write
\begin{equation}
D(\phi) = Z D_R(\phi_R)  \,\, .
\end{equation}
In terms of the renormalized quantities the classical action
(\ref{eq:classical}) reads
\begin{eqnarray}
S &=& \int_x \Bigg( \frac{1}{2} \partial_{\mu} \phi_R \partial^{\mu} \phi_R
- \frac{1}{2} m_R^2 \phi_R^2 - g^2_R \phi_R^4
\nonumber\\
&& + \frac{1}{2} \delta Z
\partial_{\mu} \phi_R \partial^{\mu} \phi_R
- \frac{1}{2} \delta m^2 \phi_R^2
- \delta g^2 \phi_R^4 \Bigg) \, .
\label{eq:counterS}
\end{eqnarray}
Similarly, one can write for the one-loop part
$\Tr\ln D^{-1} = \Tr\ln D_R^{-1}$ up to an irrelevant, temperature
independent constant. The next term of Eq.~(\ref{eq:2PIaction}) can be written
as
\begin{eqnarray}
\frac{i}{2} \Tr\, D_{0}^{-1}(\phi) D(\phi) &=&
- \frac{1}{2} \int_x \left(\square_x + m_R^2 + \delta Z_1\, \square_x
+ \delta m_1^2 \right)
D_R(x,y;\phi_R)|_{x=y}
\nonumber\\
&& - 6 \left(g^2_R + \delta g_1^2\right) \int_x
\phi_R^2(x) D_R(x,x;\phi_R)\, .
\label{eq:counter1loop}
\end{eqnarray}
Here $\delta Z_1$, $\delta m_1^2$ and $\delta g_1^2$
denote the same counterterms as introduced in (\ref{eq:introcount}),
however, approximated to the given order.
To express $\Gamma_2$ in terms of renormalized quantities it is
useful to note the identity
\begin{equation}
\Gamma_2[\phi,D(\phi)]|_{g^2}
= \Gamma_2[\phi_R,D_R(\phi_R)]|_{g^2_R + \delta g^2}
\, ,
\label{eq:counter2PI}
\end{equation}
which simply follows from the standard relation between
the number of vertices, lines and fields by counting factors of $Z$.
Therefore, one can replace in $\Gamma_2$ the bare field and
propagator by the renormalized ones if one replaces bare by renormalized
vertices as well. We emphasize that mass and wavefunction renormalization
counterterms, $\delta Z$ and $\delta m^2$, do not appear explicitly
in $\Gamma_2$. This can be understood from the fact that the
only two-particle irreducible diagrams with mass and field strength
insertions are those displayed in Fig.~\ref{fig:ct}.
The counterterms in the classical action (\ref{eq:counterS}),
in the one-loop term (\ref{eq:counter1loop}) and beyond one-loop
contained in $\Gamma_2$ have to be calculated for a given approximation
of $\Gamma_2$. Here we consider the 2PI effective action
to order $g_R^4$ with
\begin{eqnarray}
\Gamma_2[\phi_R,D_R(\phi_R)] &\!\! = \!\!&
- 3 g_R^2 \int_x D_R^2(x,x;\phi_R)
+  i 48 g_R^4
\int_{x y} \phi_R(x) D_R^3(x,y;\phi_R) \phi_R(y)
\nonumber\\
&&+  i 12 g_R^4 \int_{x y}  D_R^4(x,y;\phi_R)
- 3 \delta g_2^2
\int_x D_R^2(x,x;\phi_R)
\, ,
\label{eq:G2orderg4}
\end{eqnarray}
where the last term contains the respective coupling
counterterm at two-loop. There are no three-loop
counterterms since the divergences arising
from the three-loop contribution in (\ref{eq:G2orderg4}) are
taken into account by the lower counterterms~\cite{BIR,BBRS2}.
The coupling counterterms are displayed diagrammatically in
Fig.~\ref{fig:tadpoles}.

One first has to calculate
the solution $D_R(\phi_R)$ obtained from the stationarity
condition (\ref{eq:stationary}) for the 2PI effective action.
For this one has to impose the
same renormalization condition as for the propagator
(\ref{eq:r1}) in Fourier space:
\begin{equation}
i D_R^{-1}(p^2;\phi_R)|_{p = 0,\phi_R=0} = - m_R^2 \, ,
\label{eq:rencondD}
\end{equation}
for given finite renormalized ``four-point'' field\footnote{Below
in Eqs.~(\ref{eq:fourfieldg4}) and (\ref{eq:fourfieldchain}) we
will see that for the present approximation this
is the four-point field that solves the Bethe-Salpeter type equation
discussed in Refs.~\cite{HeesKnoll,BIR}.}
\begin{equation}
V_R(x,y;z,w) \equiv
\frac{\delta^2 i D_R^{-1}(x,y;\phi_R)}{\delta \phi_R(z)
\phi_R(w)}\Big|_{\phi_R=0} \, .
\label{eq:4field}
\end{equation}
For the above approximation we note the identity
\begin{equation}
\frac{\delta^2 \Gamma[\phi_R,D_R(\phi_R)]}
{\delta \phi_R(x) \phi_R(y)}\Big|_{\phi_R = 0}
\equiv i D_R^{-1}(x,y;\phi_R)|_{\phi_R = 0}
\label{eq:Dident}
\end{equation}
for
\begin{equation}
\delta Z = \delta Z_1 \quad ,\quad
\delta m^2 = \delta m_1^2 \quad , \quad \delta g_1^2 = \delta g_2^2 \, ,
\label{eq:renconst}
\end{equation}
such that (\ref{eq:rencondD}) for $D_R$ is trivially fulfilled.
In contrast to the exact theory, for the 2PI effective action
to order $g_R^4$ a similar identity does not connect the proper
four-vertex with $V_R$.\footnote{We
emphasize that for more general approximations the equation
(\ref{eq:Dident}) may only be valid up to higher order corrections as well.
This is a typical property of self-consistent resummations, and it does
not affect the renormalizability of the theory.
In this case the proper renormalization procedure still
involves, in particular, the conditions (\ref{eq:rencondD}) and
(\ref{eq:Vren}). For a detailed discussion of these aspects see
Ref.~\cite{BBRS2}.} Here the respective condition for the
four-point field $V_R$ in Fourier space reads
\begin{equation}
V_R(p_1,p_2,p_3)|_{p_1 = p_2 = p_3 = 0} = - 4!\, g^2_R \, .
\label{eq:Vren}
\end{equation}
Note that this has to be
the same as for the four-vertex (\ref{eq:r3}).
For the universality class of the $\phi^4$ theory there are
only two independent input parameters, which we take to
be $m_R$ and $g^2_R$, and for the exact theory
$V_R$ and the four-vertex agree.
The renormalization conditions (\ref{eq:r1})--(\ref{eq:r3}) for the propagator
and four-vertex, together with the 2PI scheme
(\ref{eq:rencondD})--(\ref{eq:Vren})
provides an efficient fixing of all the above
counterterms. In particular, it can be very
conveniently implemented numerically, which turns out to be crucial for
calculations beyond order $g_R^2$.

\begin{figure}[t]
\begin{center}
\figput{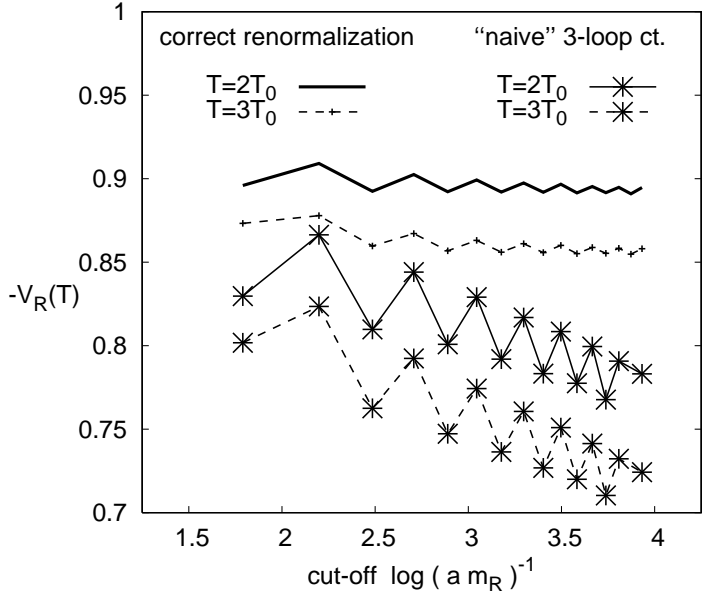}
\end{center}
\vspace*{-0.5cm}
\caption{The upper solid and dashed line show the results
for $V_R(T)$ as a function of
the logarithm of the lattice cutoff for
two temperatures $T = 2 T_0$ and $T = 3 T_0$. Here
$T_0$ is fixed by $T_0 = m_R(T_0)$ and $V_R(T_0) = - 1$.
For illustration we also show results for the same quantity
(lower curves) if the renormalization is {\em not} done properly,
by adding a ``naive'' three-loop coupling counterterm as explained in the text.
(The ``oscillation'' is a lattice artefact arising from Fourier
transformation of odd or even arrays.)}
\label{fig:climit}
\end{figure}

The apparent insensitivity\footnote{Of course,
we recover the ``triviality'' of $\phi^4$-theory
such that the four-vertex vanishes in
the continuum limit, which is discussed in Sects.~\ref{sec:analytic}
and~\ref{sec:discussion}.} of
the renormalized quantities with
respect to changes in the (lattice) regularization is
demonstrated for the temperature dependent
four-point field $V_R(T)$ in Fig.~\ref{fig:climit}.
The upper solid and dashed line show the results
for $V_R(T)$ as a function of the
logarithm of the lattice cutoff for
two temperatures $T = 2 T_0$ and $T = 3 T_0$. Here
$T_0$ is fixed by $T_0 = m_R(T_0)$ and the lattice volume is
kept constant with $L m_R(T_0)=2$.
For illustration we also show the behavior of
$V_R(T)$ if the renormalization is {\em not} done properly. This is
achieved by replacing in the three-loop contribution of (\ref{eq:G2orderg4})
$g_R^4$ by the ``bare'' coupling $(g_R^2 + \delta g_3^2)^2$ and
taking $\delta g_3^2 = \delta g_1^2$.
The induced strong logarithmic cutoff dependence of $V_R(T)$
can be observed from the respective results displayed in the
lower curves of Fig.~\ref{fig:climit}.

We emphasize that the current approximation (\ref{eq:G2orderg4})
for the 2PI effective action can only be expected to be valid for sufficiently
small $\phi_R \ll m_R/g_R$.  If the latter is not fulfilled there are
additional ${\mathcal O} (g_R^4)$ contributions at three-loop $\sim g_R^6
\phi_R^2$ and $\sim g_R^8 \phi_R^4$. The approximation should therefore not be
used to study the theory in the spontaneously broken phase or near the critical
temperature of the second-order phase transition. For studies of the latter
using the 2PI effective action see Ref.~\cite{Alford:2004jj}.
Here we will investigate how the 2PI effective action cures convergence
problems of high-temperature perturbation theory, for which the approximation
(\ref{eq:G2orderg4}) provides the correct generating functional up to order
$g_R^6$ corrections, including all
three-loop corrections in the high-temperature phase.

\subsubsection{Renormalized equations for the two- and four-point functions}
\label{ssec:renequations}

>From the 2PI effective action to order $g_R^4$ we find with
$\delta g_1^2 = \delta g_2^2$ from (\ref{eq:renconst})
for the two-point function:
\begin{eqnarray}
i D_R^{-1}(x,y;\phi_R) &\!\!=\!\!&
- \left[ (1+ \delta Z_1) \square_x + m_R^2 + \delta m_1^2 \right.\nonumber\\
&&\left. + 12 (g_R^2 + \delta g_1^2)
\left( D_R(x,x;\phi_R)+\phi_R^2(x) \right) \right] \delta(x-y)
\nonumber\\
&&
 + i 288 g_R^4 D_R^2(x,y;\phi_R)\phi_R(x)\phi_R(y)
 + i 96 g_R^4 D_R^3(x,y;\phi_R)
\, .
\nonumber\\
&&
\label{eq:invD}
\end{eqnarray}
According to (\ref{eq:Dident}) this expression coincides with the one for the
propagator
$\delta^2 \Gamma[\phi_R,D_R(\phi_R)]/\delta \phi_R(x) \delta \phi_R(y)$
at $\phi_R = 0$. It is straightforward to verify this using
\begin{equation}
\frac{\delta \Gamma[\phi_R,D_R(\phi_R)]}{\delta \phi_R(x)}
\equiv \frac{\delta \Gamma[\phi_R,D_R]}{\delta \phi_R(x)} \, ,
\end{equation}
which is valid since the variation of $D_R(\phi_R)$ with $\phi_R$ does not
contribute due to the stationarity condition (\ref{eq:stationary}).
The four-point field (\ref{eq:4field}) in this approximation is
given by
\begin{eqnarray}
V_R(x,y;z,w) &=& -24 (g_R^2 + \delta g_1^2) \delta(x-y)
\delta(x-z) \delta(x-w)
\nonumber\\
&& - 12 (g_R^2 + \delta g_1^2)
\frac{\delta^2 D_R(x,x;\phi_R)}{\delta \phi_R(z) \delta
\phi_R(w)}\Big|_{\phi_R=0} \delta(x-y)
\nonumber\\
&&+ i \frac{(24 g_R^2)^2}{2} \Bigg( \delta(x-w) \delta(y-z) + \delta(y-w)
\delta(x-z)
\nonumber\\
&&+ \frac{\delta^2 D_R(x,y;\phi_R)}{\delta \phi_R(z)
\delta \phi_R(w)}\Big|_{\phi_R=0}\Bigg) D_R^2(x,y;\phi_R=0) \, .
\label{eq:fourfieldg4}
\end{eqnarray}
Inserting the chain rule formula
\begin{equation}
\frac{\delta^2 i D_R(x,y;\phi_R)}{\delta \phi_R(z) \delta
\phi_R(w)}\Big|_{\phi_R = 0} =
- \left.
\int_{u,v} D_R(x,u;\phi_R) V_R(u,v;z,w) D_R(v,y;\phi_R)\right|_{\phi_R = 0} \,
\label{eq:fourfieldchain}
\end{equation}
into Eq.~(\ref{eq:fourfieldg4}) one arrives at the Bethe-Salpeter type
equation for $V_R$ discussed in Refs.~\cite{HeesKnoll,BIR}.
Eqs.~(\ref{eq:fourfieldg4}) and
(\ref{eq:invD}) form a closed set of equations for
the determination of the counterterms $\delta Z_1$, $\delta m_1^2$
and $\delta g_1^2$. Together with $(\ref{eq:renconst})$ one
observes that $\delta g^2$ would be undetermined from these equations
alone as employed in Refs.~\cite{HeesKnoll,BIR}. This counterterm
is determined by taking into account the equation for the physical
four-vertex, which is obtained from the above 2PI effective action as
\begin{eqnarray}
\frac{\delta^4 \Gamma[\phi_R,D_R(\phi_R)]}{\delta \phi_R(x)
\delta \phi_R(y) \delta \phi_R(z) \delta \phi_R(w)}
\Big|_{\phi_R=0} = - 24 (g_R^2 + \delta g^2)
\delta(x-y) \delta(x-z) \delta(x-w)
\nonumber\\
+ V_R(x,y;z,w) + V_R(x,z;y,w) + V_R(x,w;y,z)
\nonumber\\[0.2cm]
- 2 \left.\left( \frac{\delta i D_R^{-1}(x,y;\phi_R)}{\delta D_R(z,w)}
+ \frac{\delta i D_R^{-1}(x,z;\phi_R)}{\delta D_R(y,w)}
+ \frac{\delta i D_R^{-1}(x,w;\phi_R)}{\delta D_R(y,z)}
\right)\right|_{\phi_R=0} \, ,
\nonumber\\
\label{eq:fourvertex}
\end{eqnarray}
where from (\ref{eq:invD}) one uses the relation
\begin{eqnarray}
\left.\frac{\delta i D_R^{-1}(x,y;\phi_R)}{\delta D_R(z,w)}\right|_{\phi_R=0}
&=& - 12 (g_R^2 + \delta g_2^2) \delta(x-y) \delta(x-z) \delta(x-w)
\nonumber\\
&& + i \frac{(24 g_R^2)^2}{2} D_R^2(x,y;\phi_R=0) \delta(x-z) \delta(y-w) \, .
\end{eqnarray}
We emphasize that
the counterterm $\delta g^2$
plays a crucial role in the broken phase, since it is
always multiplied by the field~$\phi_0$ and hence it is essential
for the determination of the effective potential.
It is also required in the symmetric phase, in particular,
when one calculates the thermal coupling using Eq.~(\ref{eq:fourvertex}).

\subsection{Analytical example: 2PI two-loop order}
\label{sec:analytic}

It is instructive to consider first the 2PI effective action
to two-loop order for which much can be discussed
analytically.\footnote{Cf.~also \cite{Drummond:1997cw,Blaizot:1999ip}
and references therein.}
In this case $Z=1$ for the scalar theory and the renormalized vacuum
mass $m_R=m_R(T=0)$ of Eq.~(\ref{eq:r1}) or,
equivalently, of Eq.~(\ref{eq:rencondD}) is
given to this order by
\begin{eqnarray}
\frac{m_R^2}{g^2}
&=& 12\, \mu^{\eps}\! \int \frac{{\rm d}^d k}{(2\pi)^d}
\left( k^2 + m_R^2 \right)^{-1}
+ \frac{m^2}{g^2}
\nonumber\\
&=& - \frac{3 m_R^2}{2 \pi^2}
\left( \frac{1}{\eps} - \ln \frac{m_R}{\bar{\mu}} + \frac{1}{2} \right)
+ \frac{m^2}{g^2} \, ,
\label{eq:massgap}
\end{eqnarray}
where $m^2 = m_R^2 + \delta m_1^2$ and $g^2 = g_R^2 + \delta g_1^2$
is the zero temperature coupling $g_R = g_R(T=0)$ and we have used
(\ref{eq:renconst}).
Here we have employed dimensional regularization and evaluated the integral
in $d = 4 - \eps$ for Euclidean momenta $k$. The bare coupling in the
action (\ref{eq:classical}) has been rescaled
accordingly, $g^2 \to \mu^{\eps} g^2$, and is dimensionless;
$\bar{\mu}^2 \equiv 4 \pi e^{-\gamma_E} \mu^2$ and
$\gamma_E$ denotes Euler's constant. Below we will employ
a lattice regularization for comparison and to go beyond
two-loop order.

Similarly, the zero-temperature
four-point function resulting from Eq.~(\ref{eq:Vren}) for the
2PI effective action to order $g_R^2$ is given by
\begin{eqnarray}
g^2_R &=& g^2 - 12 g^2 g^2_R\,
\mu^{\eps}\! \int \frac{{\rm d}^d k}{(2\pi)^d}
\left( k^2 + m_R^2 \right)^{-2}
\nonumber\\
&=& g^2 - \frac{3 g^2}{2 \pi^2} g^2_R
\left( \frac{1}{\eps} - \ln \frac{m_R}{\bar{\mu}} \right) \,
\label{eq:rencoup}
\end{eqnarray}
with (\ref{eq:renconst}). We emphasize that the same equation is
obtained starting from
the renormalization condition for the proper four-vertex (\ref{eq:r3})
with $\delta g^2 = 3 \delta g_1^2$. One observes that all
counterterms are uniquely fixed by the renormalization procedure
put forward in the previous sections.

At non-zero temperature one obtains renormalized
equations for the thermal mass and coupling in terms of the vacuum parameters:
\begin{eqnarray}
m_R^2(T)&=&m_R^2 -24 g_R^2 P_0'\left(T,m_R^2(T)\right)
\nonumber\\
&&+\frac{3g^2_R}{4\pi^2}\left(m_R^2(T)\log\frac{m^2_R(T)}{m^2_R}
-m_R^2(T)+m_R^2\right)\, ,
\label{eq:gapnonzeroT}
\\
\frac{1}{V_R(T)}&=&-\frac1{4! g^2_R}-P_0''(T,m_R^2(T)) \, ,\\
g^2_R(T)&=&\frac{3}{1/g^2_R+ 4! P_0''(T,m_R^2(T))}-2g_R^2\, ,
\end{eqnarray}
where $P_0$ denotes the pressure of the free gas:
\begin{equation}
P_0(T,m^2_R(T)) = - T \int \frac{{\rm d}^3 p}{(2 \pi)^3}
\ln \left( 1 - e^{-\omega_\bp(T)/T} \right) \, .
\end{equation}
with $\omega_\bp(T) = \sqrt{\bp^2 + m^2_R(T)}$
and $P_0'(T,m_R^2(T))$ and $P_0''(T,m_R^2(T))$ denote its first and
second derivative with respect to $m_R^2(T)$.

The pressure to two-loop order is given by
\begin{eqnarray}
P_2 = P_0(T,m^2_R(T)) - \frac{1}{2} m_R^2(T)
P_0'(T,m_R^2(T))
+ \frac{m_R^4(T) - m_R^4}{128 \pi^2}
\nonumber\\
- \frac{1}{16} \left( \frac{1}{3 g^2_R} +
\frac{1}{4 \pi^2}
\right)
\left(m_R^2(T) - m_R^2 \right) m_R^2 \,.
\label{eq:pressureo2}
\end{eqnarray}
We have used these analytical formulae to check the (two-loop) numerics for
the continuum and thermodynamic limit. We have also checked, that for
the employed parameters the zero temperature limit is already reached for
$T\le m_R(T_0)/8$. We use this below to numerically estimate
the observables in vacuum.

The scalar $\phi^4$-theory in $3+1$ dimensions is non-interacting if it is
considered as a fundamental theory valid for arbitrarily high momentum scales.
This is the so-called ``triviality'' of $\phi^4$-theory~\cite{LW}. However,
if the theory is considered as a low-momentum effective theory with a
physical highest momentum then the renormalized coupling $g_R$ can be non-zero.
A non-vanishing zero-momentum four-vertex $g_R$ requires both
a highest momentum cutoff and an ``infrared cutoff'' such as
a non-zero mass term $m_R$. This can be conveniently
discussed by introducing a scale $\Lambda$ with
\begin{equation}
\frac{3}{2 \pi^2} \ln \frac{\Lambda}{\bar{\mu}}
\equiv \frac{1}{g^2} + \frac{3}{2 \pi^2 \eps} \, .
\end{equation}
In terms of this scale the renormalized coupling (four-vertex) reads
\begin{equation}
g^2_R = \frac{2 \pi^2}{3 \ln \frac{\Lambda}{m_R}} \, .
\label{eq:max}
\end{equation}
As a consequence, the zero-temperature coupling
$g^2_R$ vanishes for $m_R / \Lambda \to 0$,
i.e.~by either sending $\Lambda \to \infty$ for fixed $m_R$ or
by taking $m_R \to 0$ for fixed scale $\Lambda$.
For given $\Lambda/m_R$ the renormalized coupling
takes on its finite maximum value (\ref{eq:max}) for
a diverging $g$ due to the Landau pole in the equation
for the coupling counterterm $\delta g^2 = g^2 - g_R^2$
obtained from (\ref{eq:rencoup}) (cf.~also the
corresponding Fig.~\ref{fig:triviality}
for the 2PI three-loop solution).

The high-temperature result can be obtained with non-zero $m_R$ for the
limit $T \gg m_R$, which is conventionally dubbed the massless limit in the
literature.  To calculate $m_R(T)$ it is then
useful to rewrite the
equation (\ref{eq:gapnonzeroT}) in terms of the renormalized four-point
function
at non-zero temperature, $g_R^2(T)$, for which the limit $m_R \to 0$
can be taken directly\footnote{Note that
at the critical temperature $T_c$ of the second
order phase transition,
determined by $m_R(T_c) = 0$, the temperature dependent
coupling vanishes as well, i.e.~$g_R^2(T) \to 0$. However, the
effective coupling of the
dimensionally reduced theory $g_R^2(T_c) T_c/m_R(T_c)$ remains
non-zero and finite. We emphasize again that
in order to quantitatively describe the universal
behavior of the theory near $T_c$ requires to go beyond the present
approximation~(cf.~e.g.~the discussion using the $1/N$ expansion of
the 2PI effective action to next-to-leading
order employed in Ref.~\cite{Alford:2004jj}, and footnote \ref{foot:1}.)}
(cf.~also the discussions in Ref.~\cite{Blaizot:2003tw}).
Here the two-loop pressure (\ref{eq:pressureo2})
in the limit $m_R \to 0$ becomes
\begin{eqnarray}
P_2(T) = P_0(T) + \frac{1}{2} m_R^2(T) P_0'(T,m_R^2(T))
+ \frac{m_R^4(T)}{128 \pi^2} \, ,
\label{eq:pressureo2massless}
\end{eqnarray}
using
\begin{equation}
\lim_{m_R \to 0} \frac{m_R^2}{g_R^2}
= \lim_{m_R \to 0} \frac{3}{2 \pi^2} m_R^2 \ln \frac{\Lambda}{m_R} = 0 \, .
\end{equation}

It is instructive to compare at this point to perturbation theory
since its characteristic problems are apparent already at low
orders. In Sec.~\ref{sec:numerics_nlo}
we will discuss aspects of the higher order
behavior. To make link with other schemes
we note that the $\overline{\rm MS}$ renormalized
``running'' $V_R(\bar{\mu})$ is given by
\begin{equation}
-\frac{4!}{V_R(\bar{\mu})} \equiv \frac{1}{g^2_R} + \frac{3}{2 \pi^2}
\ln \frac{m_R}{\bar{\mu}}
= \frac{1}{g^2} + \frac{3}{2 \pi^2 \eps}\, ,
\end{equation}
which can be written explicitly as
\begin{eqnarray}
-\frac{1}{4!}V_R(\bar{\mu})
= \frac{g^2_R}{1 + \frac{3 g^2_R}{2 \pi^2}
\ln \frac{m_R}{\bar{\mu}}} \, .
\end{eqnarray}
One observes that $V_R(\bar{\mu})$ corresponds to
the zero-momentum four-vertex $g_R^2$ for the choice of $\bar{\mu} = m_R$.

In perturbation theory the massless or high-temperature limit has been
particularly extensively studied in the literature.
Up to order $g_R^4$ corrections
the weak coupling expansion of the pressure for $T \gg m_R$
can be obtained starting
from (\ref{eq:pressureo2}) or (\ref{eq:pressureo2massless})
using the high-temperature expansion
of the gap equation (\ref{eq:gapnonzeroT}):
\begin{equation}
m_R^2(T) \simeq g_R^2 T^2 - \frac{3}{\pi} g_R^2 T m_R(T) \, ,
\end{equation}
where we neglect perturbative terms of order $g_R^4$ and
higher. The first term comes from
$12 g_R^2 \int \frac{{\rm d}^3 p}{(2\pi)^3} n(\bp)/|\bp| = g_R^2 T^2$
and the second term employs
\begin{eqnarray}
\int \frac{{\rm d}^3 p}{(2\pi)^3} \left(\frac{n(\omega_\bp)}{\omega_\bp}
- \frac{n(\bp)}{|\bp|} \right) \, \simeq\,  T
\int \frac{{\rm d}^3 p}{(2\pi)^3} \left( \frac{1}{\bp^2 + m_R^2(T)}
- \frac{1}{\bp^2} \right) \, =\, - \frac{1}{4 \pi} m_R(T) T \, ,
\nonumber
\end{eqnarray}
with $n(\omega_\bp) = (\exp(\omega_\bp/T) - 1)^{-1}$ and where it was used that
the momentum integral is dominated by momenta $\sim g_R T$ with
$n(\omega_\bp) \simeq T/\omega_\bp$. Using these approximations in the
2PI two-loop pressure (\ref{eq:pressureo2massless}),
the high-temperature perturbative result is obtained as
\begin{equation}
\frac{P_{\rm pert} (T)}{T^4} \simeq
\frac{\pi^2 }{90} - \frac{g_R^2}{48} + \frac{g_R^3 }{12 \pi} \, ,
\label{eq:pressuremzero}
\end{equation}
up to terms of order $g_R^4$~\cite{Blaizot:2003tw}.
The entropy and energy density, respectively, are then given by
${\mathcal S}_{\rm pert}(T) \simeq 4 P_{\rm pert}(T)/T$ and
${\mathcal E}_{\rm pert}(T) \simeq 3 P_{\rm pert}(T)$.
The perturbative results to
order $g_R^2$ and $g_R^3$ are displayed in Fig.~\ref{fig:pressure}
along with the 2PI two-loop result.
The alternating behavior and poor convergence of the perturbative
series observed from the low orders in (\ref{eq:pressuremzero}) manifest itself
also at higher orders of $g_R$~\cite{fiveloop}. This problem
of perturbation theory is not specific to the limit $T \gg m_R$ but
can also be observed for lower $T/m_R$. For instance, for
$T \ll m_R$ the perturbative
(positive) order $g_R^4$ contribution to the pressure
is found to dominate the (negative) order $g_R^2$ contribution
for not too small coupling~\cite{fiveloop}.

\section{Numerical calculation to three-loop order}
\label{sec:numerics}

In order to obtain the results for the 2PI effective action
to three-loop order without further
approximations, we use numerical (lattice) techniques. The employed
hypercubic lattice discretization provides a regularization scheme.
It turns out to be convenient to carry out the calculation
using the unrenormalized propagator $D$ and four-point field $V$,
where $V = Z^{-2} V_R$ defined with Eq.~(\ref{eq:4field}).
On the lattice we consider Euclidean space-time and in
Fourier space we denote the Euclidean two-point field by
$\bar{D}_k$ and four-point field by $\bar{V}_k \equiv \bar{V}(k,-k,0,0)$,
without loss of generality.
Here $\bar{D}_{k=0}^{-1} = Z^{-1}m_R^2$, $d \bar{D}_k^{-1}/d k^2|_{k=0}
= Z^{-1}$ and
$\bar{V}_{k=0} = 24 Z^{-2}g_R^2$ are the Euclidean equivalents of the
above renormalization conditions. Following the procedure of
Sec.~\ref{sec:renormalization}, for the renormalization
one needs with (\ref{eq:renconst}) to know
$g^2$, $\delta g^2$, $m^2$ and $Z$ which are functions of the
lattice cutoff. The errors arising from subtracting cutoff dependent
quantities such as the quadratic mass contribution of the setting sun diagram
remain under control, since the numerical value of the unrenormalized
quantities as well as the physical values simultaneously fit into a double
precision variable for the employed cutoff values.

On the lattice there is only a subgroup of the rotation symmetry generated
by the permutations of $p_x,p_y,p_z$ and the reflections $p_x
\leftrightarrow -p_x$ etc., which entails a reduction of independent
lattice sites by a factor of 48. For three space dimensions, using
periodic boundary conditions, a lattice of typical linear size $32 a$
requires $N_s=969$ independent sites. In the fourth
dimension we use a different lattice spacing ($a_t$) and lattice
size (typically $N_t = 128$) so the rotation symmetry cannot be extended.
We implement two routines that make use of the symmetry
features in order to have acceptable performance, the addup function
and a four dimensional fast Fourier transformation defined as:
\begin{eqnarray}
\addup(\bar{H})&=&\frac1{L_4}\sum_k \bar{H}_k \, ,\label{eq:addup}\\
\fft(\tilde{H})_k&=&a^3a_t\sum_x \tilde{H}_x e^{ikx} \, ,\\
\invfft(\bar{H})_x&=&\frac1{L_4}\sum_k \bar{H}_k e^{-ikx} \, ,\label{eq:invfft}
\end{eqnarray}
where $L_4=N^3N_ta^3a_t$ and $\bar{H}$ is a given lattice field in
Fourier space.\footnote{
  For $x=0$ the values of Eqs.~(\ref{eq:addup}) and (\ref{eq:invfft}) are
  identical. Both routines are implemented separately to reduce
  computational costs.}
Fields with a tilde ($\tilde{H}$) are in coordinate
space. Here $\Sigma_k$ (and $\Sigma_x$) denotes a summation over all lattice
four-momenta (or coordinates) respecting the appropriate symmetry weights.

\subsection{Two-loop solution}
\label{sec:numerics_lo}

We first consider the two-loop lattice calculation, which can be
compared with the analytic discussion of Sec.~\ref{sec:analytic}.
In this case $\bar{D}_k^{-1} = k^2 + m_R^2$ and
the respective coupling and mass equations on the lattice read
\begin{eqnarray}
1/g^2 &=& 1/g_R^2 - 12\, \addup(\bar{D}^2) \, ,
\\
m^2 &=& m_R^2 - 12 g^2\, \addup(\bar{D}) \, .
\end{eqnarray}
We solve for $g^2$ and $m^2$ for given renormalized parameters
as a function of the lattice cutoff. With these bare parameters
we calculate the pressure, thermal mass
and coupling at various temperatures, using the same lattice
spacing. The results are given below along with the order $g_R^4$
results. The variation of the temperature is implemented by changing the
lattice size according to $T=1/(N_t a_t)$.

The pressure is calculated by evaluating the 2PI effective action
to this order using Eq.~(\ref{eq:relpg}):
\begin{equation}
P_2 = \addup\left[ \frac12\log \bar{D} + 3 g^2\bar{D} \addup \bar{D}\right]
\, .
\end{equation}
This expression suffers from a temperature independent quartic divergence.
To subtract this, we always measure pressure differences, by calculating
the pressure at zero temperature as well. We carry out
the subtraction before calling the spatial part of the overall \addup
function. The sum over the fourth dimension has to be done
beforehand, since the respective lattice sizes are not the same
for different temperatures.

\subsection{Three-loop solution}
\label{sec:numerics_nlo}

We obtain the renormalized physical results in two steps:
  At zero temperature or a given temperature $T_0$ we numerically
  solve the equation (\ref{eq:invD}) for
  the two-point field and the equation (\ref{eq:fourfieldg4}) for the
  four-point field with the conditions (\ref{eq:rencondD})
  and (\ref{eq:Vren}) simultaneously.
  This way we obtain the counterterms $\delta g_1^2$, $\delta Z_1$
  and $\delta m_1$, using (\ref{eq:renconst}). The counterterm for
  the coupling in the classical action,
  $\delta g^2$, can then be obtained from Eq.~(\ref{eq:fourvertex}).
  At a different temperature $T$ we use the counterterms obtained in the first
  step to evaluate physical quantities. We solve first Eq.~(\ref{eq:invD})
  and get the thermal propagator, the thermal mass and the pressure.
  Then by solving the corresponding equation (\ref{eq:fourfieldg4}) we
  obtain the four-point function $\bar{V}_k(T)$. Finally,
  the thermal coupling is obtained from Eq.~(\ref{eq:fourvertex}) using the
  previously obtained value of~$\delta g^2$.

In the following we describe the numerical implementation of the
simultaneous iterative solution of the propagator
and coupling equation in more detail.
The renormalized mass, $m_R$, is extracted from the
low-momentum form of the inverse propagator
$\bar{D}_k^{-1} = Z^{-1}\left(k^2+m_R^2+{\cal O}(k^4)\right)$.
The renormalized coupling, $g_R^2$ is obtained from the
respective vertex equation for $\bar{V}_k$ to this order:
\begin{eqnarray}
\frac{\bar{V}_k}{24} &=& g^2  - \frac{g^2}{2 L_4} \sum_q \bar{D}_q^2 \bar{V}_q
- 24 Z^{-4}g_R^4 \bar{F}_k
+ \frac{12 Z^{-4}g_R^4}{L_4}\sum_q \bar{F}_{k+q} \bar{D}_q^2 \bar{V}_q \, ,
\end{eqnarray}
with $\bar{F}_k \equiv L_4^{-1} \Sigma_q \bar{D}_q \bar{D}_{k-q}$. We note that
both the integrand and the kernel $\bar{F}_k$ from the sunset diagram
can be calculated by convolution, which is simple if the
\fft and \invfft routines are provided.
The numerical implementation of the vertex equation can be summarized
as follows:

\hspace*{2cm}\parbox{12cm}{
\noindent$\tilde{F}_x =\left( \invfft(\bar{D}) \right)^2_x$\\
$\bar{F}_k = \fft(\tilde F)_k $\\
$\texttt{begin loop}~\textrm{(iterations)}$\\
\hspace*{1cm}$\tilde H_x = \invfft (\bar{D}^2 \bar{V})_x$\\
\hspace*{1cm}$I_k = \fft(\tilde F \tilde H)_k$\\
\hspace*{1cm}$\bar{V}^{\rm new}_k/24 = g^2 - g^2 {\tilde H}_{x=0}/2
               - 24 g_R^4 Z^{-4} \bar{F}_k + 12 g_R^4 Z^{-4} I_k$\\
$\texttt{end loop}$
}

\begin{figure}[t]
\begin{center}
\figput{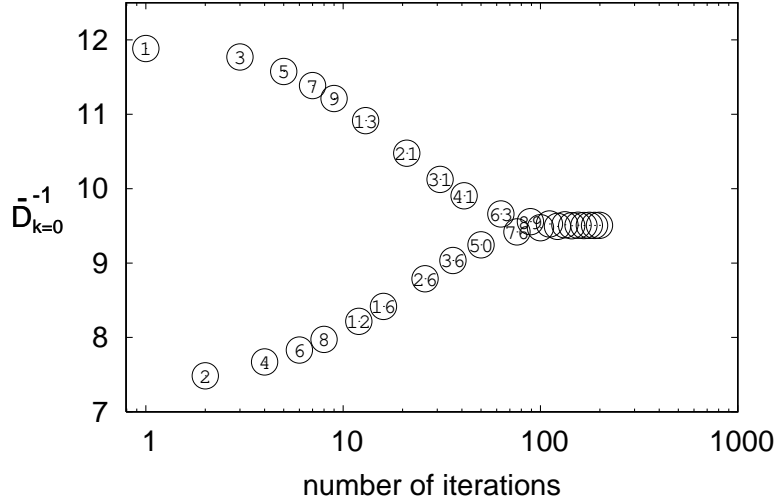}
\end{center}
\vspace*{-0.5cm}
\caption{Shown is the behavior of the naive iterative solution of
the equation for the propagator. Here each number corresponds to
the value of $\bar{D}_{k=0}^{-1}$ after the respective number of iterations,
starting from the classical propagator $\bar{D}_{0,k}$. The temperature
is $T= 4 m_R$ in units of the zero-temperature mass $m_R$.
One observes convergence only after of order 100 iterations.
Iterative solution procedures can nevertheless be very efficiently
applied with slight modifications (with a convergence
factor of 0.5 convergence is achieved after 14 steps,
cf.~the discussion in the text).}
\label{fig:conv}
\end{figure}

\noindent
Similarly, the propagator equation can be implemented by convolutions:

\hspace*{2cm}\parbox{12cm}{
\noindent$\texttt{begin loop}~\textrm{(iterations)}$\\
\hspace*{1cm}$\tilde D=\invfft(\bar{D})$\\
\hspace*{1cm}$\bar{D}^{-1, {\rm new}}_k = \bar{D}_{0,k}^{-1}
+ 12 g^2 \addup(\bar{D})
- 96 g_R^4 Z^{-4}\fft ( {\tilde D}^3)_k$\\
$\texttt{end loop}$
}

\noindent
with $\bar{D}_{0,k}^{-1}=k^2+m^2$.
We emphasize that the simple iterations do not converge well. The iterated
propagator and four-point function oscillates
around a physically senseful value
and sometimes this oscillation is damped slower than the round-off errors
accumulate. The alternating behavior at each order in the
iteration is exemplified for the zero-momentum propagator
in Fig.~\ref{fig:conv}.
A more efficient iterative solution procedure involves
a convergence factor, which avoids/damps the alternating behavior:
\begin{equation}
\bar{D}_k^{-1, {\rm update}}= \alpha \bar{D}_k^{-1, {\rm new}}
+ (1-\alpha) \bar{D}_k^{-1} \, .
\end{equation}
For a choice of $\alpha=0.1\dots 0.5$ convergence was typically
achieved after $20 \dots 50$ iterations by matching the exit criterium.
This criterium is based on the sufficient smallness
of the absolute change in $\bar{D}$ or $\bar{V}$ at all momenta.

We note that the naive iterative solution of the equation for the
propagator (with $\alpha=1$)
reflects some aspects of a perturbative calculation.
Starting from the classical propagator,
each iteration adds a higher order contribution to $\bar{D}$.
At low orders the origin of the problematic
oscillating behavior of a perturbative series in $g_R$
can be nicely observed from Fig.~\ref{fig:conv}.
Though higher orders in the iteration
do not take into account all respective perturbative contributions,
it is interesting that a convergence is obtained only
after of order $100$ (!) iterations.

When fixing the renormalized theory we implement an outer loop for
repeating the subsequent solutions of the vertex and the propagator
equation. The bare coupling is tuned within the vertex equation iterations and
the bare mass is adjusted after each iteration in the propagator equation.
Then, using Eq.~(\ref{eq:fourvertex}), simple algebra gives $\delta g^2$.
Calculating the thermodynamics at a given temperature using the previously
obtained bare parameters is much simpler: first we solve the equation
for the two-point field without any tuning of the parameters, then we obtain
the thermal coupling from the vertex equation and no outer loop is required
here.
In addition to the two-loop contribution to the pressure,
$P_2$, there is an additional contribution from the basketball
diagram at three-loop order such that the pressure is given by
\begin{eqnarray}
P_3 &=& \addup\left[ \frac12\log \bar{D} + 3 g^2\bar{D} \addup \bar{D}
- 36 g_R^4 Z^{-4} \bar{D} \fft {\tilde D}^3 \right]\, ,
\end{eqnarray}
with $\tilde D = \invfft \bar{D}$. The subtraction of the divergences proceeds
exactly in the same way as in the two-loop case: we are subtracting
spatial lattice fields and we can only then carry out the spatial integral.

Though this analysis is carried out using a lattice discretization, 
there is no principle obstacle exploiting the full rotation 
invariance of the continuum equations. This will be 
important in order to discuss the high-temperature behavior
in applications of these techniques to more complex 
theories such as QCD. 

\section{Discussion}
\label{sec:discussion}

On the left of
Fig.~\ref{fig:pressure} the two- and three-loop results for the pressure are
shown  as a function of the renormalized coupling $g_R$.
For the employed high temperature $T = 2 T_0$ with $T_0=m_R(T_0)$
we observe that the three-loop correction to the pressure is rather small.
We note that here the three-loop thermal mass is larger than the two-loop mass,
which drives the three-loop pressure below the two-loop value.
\begin{figure}[t]
\begin{center}
\centerline{
\parbox{7.9cm}{\hbox{\epsfig{file=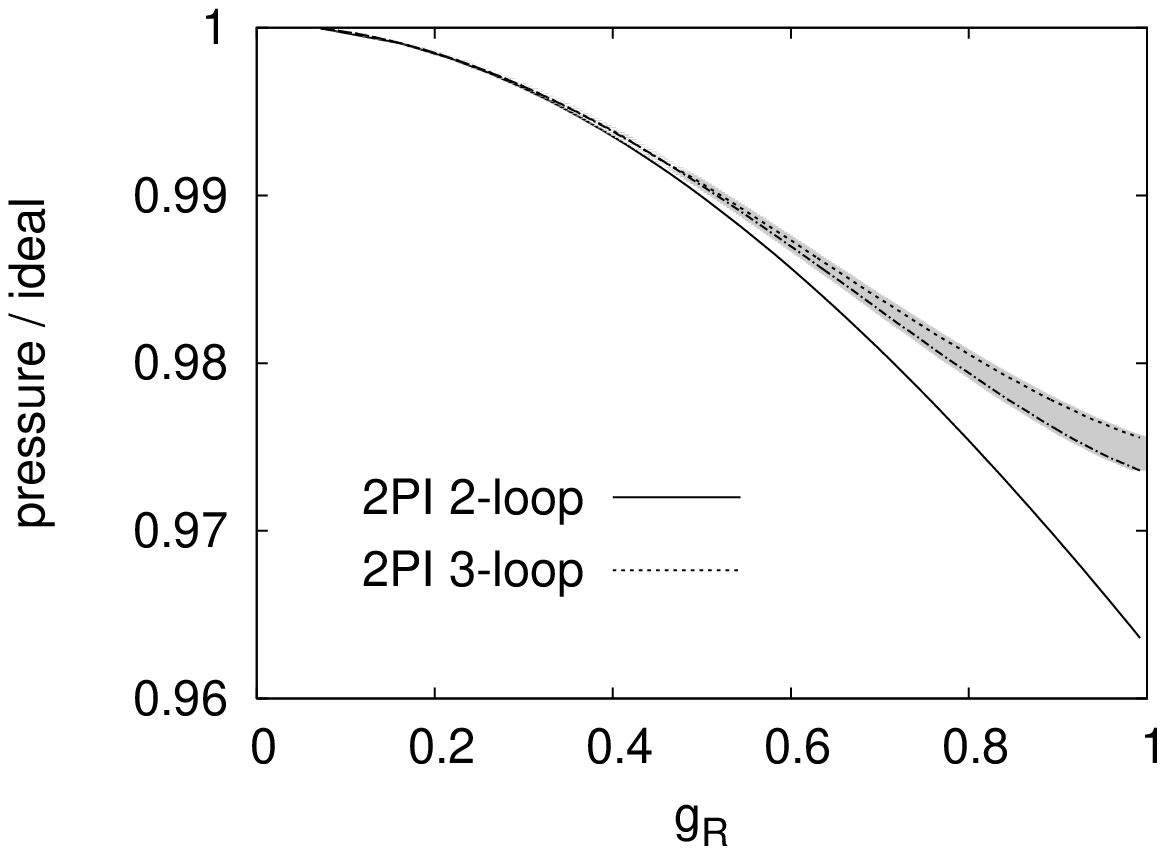,width=7.9cm}}}
\hspace*{-0.3cm}
\parbox{7.9cm}{\hbox{\epsfig{file=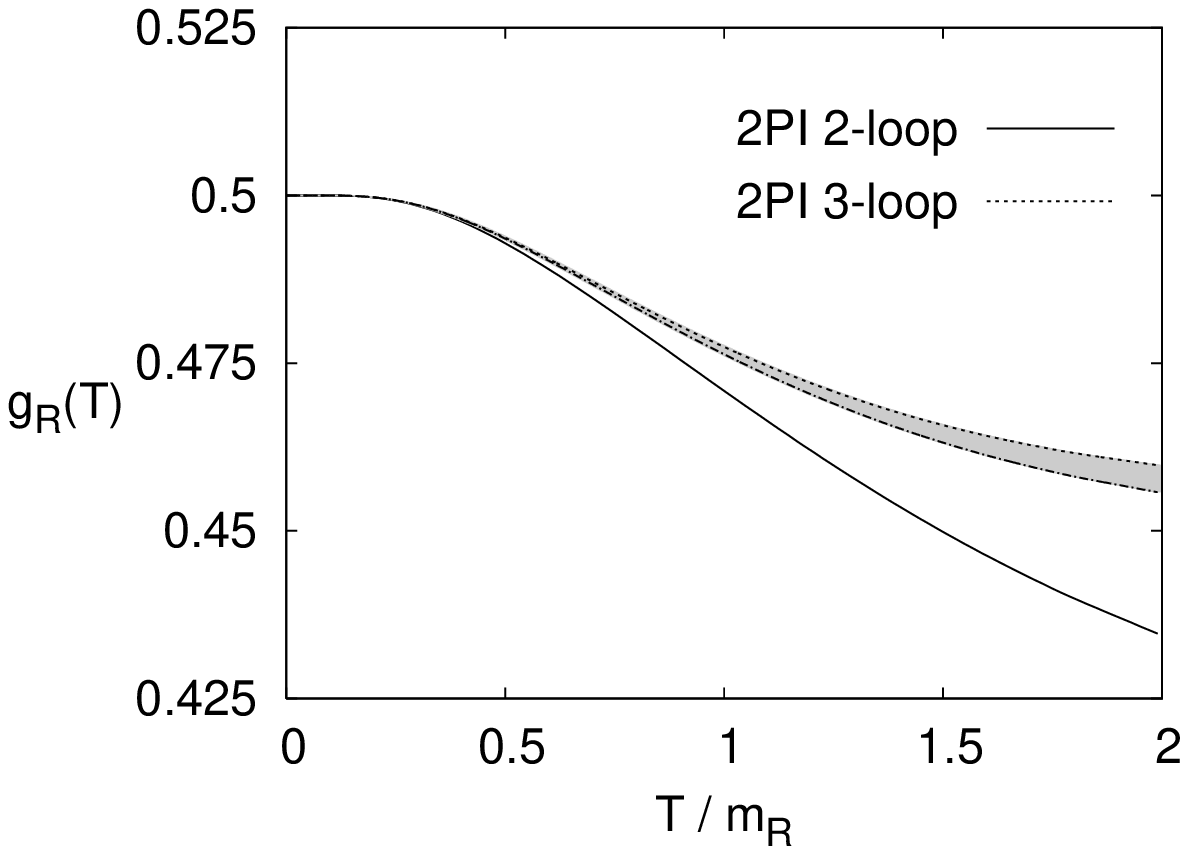,width=7.9cm}}}
}
\end{center}
\vspace*{-0.5cm}
\caption{The left figure shows the results for pressure at a lower
temperature ($T = m_R(0)$) than Fig.~\ref{fig:pressure} as a function of
the renormalized zero-temperature coupling. The right figure
shows the renormalized thermal coupling as a function of temperature.
As for Fig.~\ref{fig:pressure} results are given employing two different
renormalization scales. The shaded band indicates the renormalization scale
dependence appearing at three-loop order.
}
\label{fig:vacuumfix}
\end{figure}

In Fig.~\ref{fig:vacuumfix} we show results at a lower temperature, which
is taken to be equal to the vacuum mass, i.e.~$T = m_R$.
We plot the pressure as a function of the vacuum coupling.
This set of parameters has the feature that the
thermal masses are within 1\% equal in the two approximations for any
coupling. In this case the three-loop curve is above the two-loop
curve. Note that the two-loop contribution from the
2PI effective action is negative, while the
three-loop correction is positive. This can be directly
observed from the 2PI effective action before evaluation
at the stationary point (\ref{eq:stationary}) for which
$i \Gamma_2^{\rm 2loop}[\phi=0,D]\, T/L_3
= - 3 g^2 (\addup \bar{D})^2$ and
$i \Gamma_2^{\rm 3loop}[\phi=0,D]\, T/L_3
= 12 (g_R/Z)^4 \addup (\bar{D} \fft {\tilde D}^3)$.
As a consequence, if the propagators do not change
much from two-loop to three-loop order then
the pressure will always be increased by the
three-loop contribution.
On the right of Fig.~\ref{fig:vacuumfix} we also show the
renormalized coupling as a function of temperature
normalized to $m_R$.

For the exact theory the choice of a temperature scale for renormalization is
irrelevant.  However, for a given approximation the possible renormalization
scale dependence of the results can be used as a check.
The two-loop result is manifestly renormalization scale independent
(cf.~also Sec.~\ref{sec:analytic}).
The three-loop results of Figs.~\ref{fig:pressure} and
\ref{fig:vacuumfix} have been calculated
first from renormalization at zero temperature and second from
renormalization at a non-zero temperature $T=m_R$. The 
high-temperature renormalization conditions are chosen such that
the results agree at zero temperature. At three loop this check
is nontrivial because of the coupling dependence in Eq.~(\ref{eq:G2orderg4})
with $g_R = g_R(T=0)$ for the first calculation and
$g_R = g_R(T=m_R)$ for the second one.
The variation of the thermal results of the two models give an idea about
the scale dependence, which is indicated by the shaded bands
in Fig.~\ref{fig:vacuumfix}.
A similar analysis employing different renormalization scales
has been also performed for the 2PI three-loop results displayed
on the left of Fig.~\ref{fig:pressure}. However, the difference
between the results was hardly distinguishable in this
high-temperature case. In contrast, the severe scale dependence
of the perturbative calculations is well documented in the
literature~\cite{Blaizot:2003tw}.

\begin{figure}[t]
\begin{center}
\epsfig{file=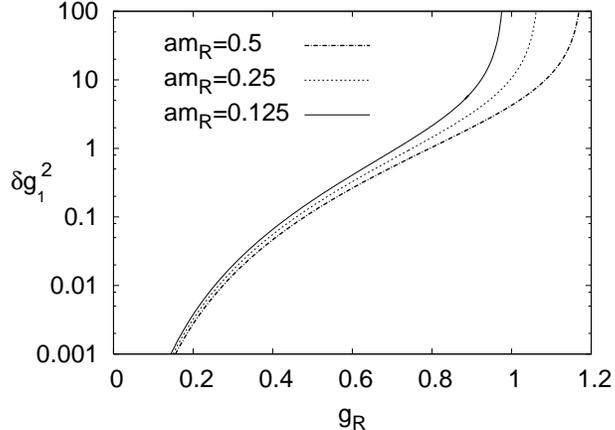,width=8.5cm}
\end{center}
\vspace*{-0.5cm}
\caption{Physical coupling ranges for different lattice spacings
employing the three-loop 2PI effective action. Here we show
the coupling counterterms $\delta g_1^2 = \delta g_2^2$.
The counterterms diverge at a certain value
of the physical coupling.
}
\label{fig:triviality}
\end{figure}

Compared to the perturbative expansions for pressure
(cf.~Fig.~\ref{fig:pressure} right), our results indicate a
substantially improved convergence for the 2PI expansions. This
holds even for rather strong couplings. In our lattice calculations
we approach the Landau pole with our momentum cutoff, i.e.\ we
explore almost the full range of couplings principally available.
The range of renormalized couplings for various lattice
regularizations is shown in Fig.~\ref{fig:triviality}. For any
momentum cutoff there is a highest value of $g_R$ for which there
exists a bare coupling $g^2 = g_R^2 + \delta g^2$. (Cf.~also the
discussion for the two-loop effective action in
Sec.~\ref{sec:analytic}.) Concerning the lattice discretization one
gains precision by reducing the lattice spacing, however, the range
of physical couplings that can be realized shrinks. From the
comparison to the two-loop analytic formulae, and from three-loop
calculations with a series of small lattice spacings, we infer that
one should use $a m_R(T_0) \le 0.125$ (for $T\le 2 m_R(T_0)$, see
Fig.~\ref{fig:climit}). How this limits the available couplings to
three-loop order is shown in Fig.~\ref{fig:triviality}.

Very similar techniques as those employed here can be straightforwardly
used for more complicated systems, as for instance fermionic or Yukawa
theories (cf.~also Ref.~\cite{Berges:2002wr}). A more ambitious generalization
is the application of these techniques to gauge field theories. While
linear symmetries as realized in QED can be treated along similar lines, the
generalization to nonabelian gauge theories is technically more
involved and needs to be further investigated \cite{BBRS2}.\\

\noindent
{\bf Acknowledgements.} We thank Jean-Paul Blaizot, Holger Gies,
Edmond Iancu and Toni Rebhan for fruitful collaborations/discussions
on these topics.\\

\end{document}